\documentclass[12pt,preprint]{aastex}

\def\apj#1#2#3#4#5{\bibitem[#1]{#2}#3, { Ap. J.}, {#4}, #5}

\def\science#1#2#3#4#5{\bibitem[#1]{#2}#3, Science, #4, #5}

\def\spie#1#2#3#4#5{\bibitem[#1]{#2}#3, SPIE, #4, #5}

\begin{document}

\title{Sensitivity of the Astrometric Technique in 
Detecting Outer Planets}

\author{J. A.  Eisner \& S. R. Kulkarni}
\affil{Palomar Observatory 105-24, California Institute of
Technology, Pasadena, CA 91125 \\ jae@astro.caltech.edu, srk@astro.caltech.edu}

\begin{abstract}
With the advent of optical interferometers that will be coming online in 
the next decade, radial velocity searches for extra-solar planets will be 
complemented by high angular resolution astrometric measurements. In this 
paper, we explore the sensitivity of the astrometric technique, and develop 
an analytical understanding of the sensitivity 
in the regime where the orbital period is longer than the time-baseline 
of the survey. As in an earlier paper that dealt with the radial velocity 
technique, 
our analysis of the astrometric technique utilizes both the orbital 
amplitude and phase.
\end{abstract}

\section{Introduction}
Radial velocity surveys of nearby stars have produced spectacular
discoveries of extra-solar planets over the past two 
decades (see Marcy, Cochran \& Mayor 2000). As these searches continue
into the next decade, they will be supplemented by high angular
resolution searches for the astrometric signatures of planetary orbits
\citep[{e.g.}][]{DU99,VANBELLE+98,HORNER+99}.
While radial velocity (RV)
technique works by detecting the reflex acceleration of the star
due to the orbital motion of the companion, astrometric technique
detects the apparent reflex motion of the star on the sky.  
In a previous paper \citep[hereafter EK2001]{EK2001},
we examined the RV technique
in detail, paying particular attention to the regime where the
time-baseline of the observations is shorter than the orbital period
of the planet.  
This regime is interesting because while the longest baseline
of current RV surveys is $\sim 12$ years, one expects giant
planets to form in the colder regions of the proto-planetary
nebula, and thus we expect such objects to possess periods of
many years to centuries \citep{BOSS95}.  
While RV technique is most sensitive to short-period, close-in planets, 
astrometric technique obtains optimal sensitivity to long-period massive 
planets expected from current theories. Moreover, 
in EK2001, we demonstrated that one can achieve
a significant improvement in sensitivity (over current methods)
if the orbital amplitude
{\em and phase} are included in the analysis.

In this paper, we develop an analytical understanding of
the sensitivity of astrometric technique
in this so-called ``long-period regime'', and address the issues of
detection and detection efficiency.  To the best of our knowledge,
the last authors who dealt with the sensitivity of the astrometric
method were Black \& Scargle (1982), who used a periodogram approach 
\citep{BS82}.
As in EK2001, we will
argue below that a least-squares approach is preferable to the
periodogram.  Essentially, the least-squares approach
(in contrast to the periodogram method) offers the most general
approach, and requires no modification in the long-period regime.

The plan of the paper is as follows.  First, we address the case of
edge-on orbits in \S \ref{sec:edge}.  In \S \ref{sec:basic},
we summarize the basic equations and describe the least-squares
approach in \S \ref{sec:lsq}.  In \S \ref{sec:typeI} we provide analytical
estimates for obtaining a detection in the absence of any signal,
and in \S \ref{sec:typeII} we 
obtain estimates for minimum detectable signals in the
presence of noise.  We repeat this analysis for the
case of face-on orbits in \S \ref{sec:face} and for orbits with general
inclination angles in \S \ref{sec:45}.  We
conclude in \S \ref{sec:concls}.

\section{Edge-On Orbits \label{sec:edge}}

\subsection{Basic Equations \label{sec:basic}}
We will assume circular orbits throughout this discussion,
and we will neglect the effect of annual parallax.  
The astrometric signature of an edge-on circular orbit is given by
\begin{equation}
\theta(t) = {\cal A} \sin \left( \frac{2\pi t}{\tau} +\phi\right)
+ \lambda t + \mu,
\label{eq:astrom-orbit}
\end{equation}
where $t$ is the time; $\lambda$ and $\mu$ are the proper motion and position 
offset of the planetary system, respectively; and
\begin{equation}
{\cal A} = \frac{M_p}{D}
\left(\frac{G \tau^2}{4\pi^2 M_{\ast}^2} \right) ^{\frac{1}{3}}. 
\label{eq:astrom-amplitude}
\end{equation}
Here, $D$ is the distance to the system, 
$M_{\ast}$ is the mass of the star, and $M_p$ is the mass of the
planet. 
Thus, we can express the sensitivity (defined as the minimum-mass planet
that can be detected) of the  astrometric technique in terms
of the amplitude of the astrometric orbital signature:
\begin{equation}
M_p = {D {\cal A}} \left(\frac{4\pi^2 M_{\ast}^2}{G \tau^2}\right)
^{\frac{1}{3}} .
\label{eq:sensitivity}
\end{equation}
However, it is more difficult to identify planets with long periods than 
Equation \ref{eq:sensitivity} might suggest.  In the so-called
``long-period regime'', the total baseline of the observations is
less than the orbital period of the planet, and thus we only observe
some fraction of the full orbital amplitude.
Thus the sensitivity is expected to critically depend
on the orbital phase.  As we will show explicitly below, 
astrometric technique (applied to an edge-on orbit) will be the
most sensitive when the orbital curvature is
maximal ({i.e.} $\phi = [n+1/2] \pi$).  In contrast, when the orbital
phase is close to $n\pi$ ({i.e.} a pure sine wave), the lack of 
orbital curvature makes it difficult to distinguish the
signal from the unknown proper motion of the system.

\subsection{Least Squares Fitting of Sinusoids \label{sec:lsq}}
The signal analysis for the astrometric technique consists
of fitting the observations to the model specified in
Equation \ref{eq:astrom-orbit}.
As noted by several authors \citep[{e.g.}][]{SCARGLE82,NA98,EK2001} 
the most optimal fitting is obtained by using the
technique of least-squares.  First, we derive a linear model
equation from Equation \ref{eq:astrom-orbit}:
\begin{equation}
\theta(t) = {\cal A}_c \cos(\omega t) + {\cal A}_s \sin(\omega t) + \lambda t + \mu.
\label{eq:lsq-model1}
\end{equation}
Here, ${\cal A}_c = {\cal A} \sin \phi$, ${\cal A}_s = {\cal A} \cos \phi$.
Black \& Scargle (1982) estimated $\mu$ and $\lambda$ first and
then applied periodogram analysis to the residual data ({i.e.} with 
contributions from $\mu$ and $\lambda$ subtracted).  However,
as discussed in EK2001, it is necessary to determine all of
the parameters simultaneously, especially in the long-period regime.

Using Equation \ref{eq:lsq-model1} as
our physical model, we perform the following analysis
to search for a signal in a data set.  The parameters of the data
set are as follows:  the duration of the survey is $T_0$,
$\theta'(t_j)$ is the measured astrometric offset at epoch $t_j$,
and $n_0$ is the number of measured epochs.  With no loss
of generality, we let the time go from $t=-T_0/2$ to $T_0/2$.
To find the four unknowns, ${\cal A}_c$, ${\cal A}_s$, $\lambda$, and $\mu$,
we minimize
\begin{equation}
X^2 = \sum_{j=0}^{n_0-1} [\theta'(t_j) - {\cal A}_c \cos(\omega t_j)
-{\cal A}_s \sin(\omega t_j) - \lambda t_j - \mu]^2
\label{eq:x2}
\end{equation}
with respect to ${\cal A}_c$, ${\cal A}_s$, $\lambda$, and $\mu$.  Here,
$\omega = 2\pi/\tau$, where $\tau$ is the orbital period.
This yields a matrix equation for the four unknowns, which must
be solved numerically in the general case.  However, in the
short-period regime where we observe many orbital cycles,
many of the summations will average to zero, and this matrix
equation will become approximately diagonal.  In this
case, the fit parameters are given by
\begin{equation}
{\cal A}_c = \frac{2}{n_0} \sum_{j=0}^{n_0-1} \theta'(t_j) \cos(\omega t_j),
\label{eq:ac-short}
\end{equation}
\begin{equation}
{\cal A}_s = \frac{2}{n_0} \sum_{j=0}^{n_0-1} \theta'(t_j) \sin(\omega t_j),
\label{eq:as-short}
\end{equation}
\begin{equation}
\lambda = \frac{\sum_{j=0}^{n_0-1} \theta'(t_j) t_j}{\sum_{j=0}^{n_0-1} t_j^2},
\label{eq:lambda-short}
\end{equation}
\begin{equation}
\mu = \frac{1}{n_0} \sum_{j=0}^{n_0-1} \theta'(t_j).
\label{eq:mu-short}
\end{equation}

\subsection{Type I Errors \label{sec:typeI}}
Type I errors describe the probability that a high amplitude
will be obtained even when no signal is present in the data.
We assess the statistics of Type I errors, assuming that our
data set consists of Gaussian noise with a mean of zero and
a standard deviation of $\sigma_0$.

\subsubsection{Short-Period Regime \label{sec:tI-short}}
From Equations \ref{eq:ac-short}--\ref{eq:mu-short}, we see
that ${\cal A}_c$, ${\cal A}_s$, $\lambda$, and $\mu$ are sums
of Gaussian variables, and thus by the Gaussian addition theorem,
the four derived parameters must also be Gaussian.  Specifically,
${\cal A}_c$ and ${\cal A}_s$ obey a Gaussian distribution with a mean of zero
and a variance of
\begin{equation}
\sigma^2 = \frac{2 \sigma_0^2}{n_0},
\label{eq:sig}
\end{equation}
where $n_0$ is the number of measurements.  Denoting by $A_{1s}$ the
value of $\vert {\cal A}_c \vert$ (or $\vert {\cal A}_s \vert$) that is
exceeded in 1\% of cases, we note that
\begin{equation}
A_{1s} = 2.61 \sigma = 3.69 \: n_0^{-1/2} \sigma_0.
\label{eq:a1s}
\end{equation}

From Equation \ref{eq:mu-short}, we see that $\mu$ obeys a
Gaussian distribution with zero mean and a variance of
$\sigma_0^2/n_0$.  Denoting by $\mu_{1s}$ the value of $\vert \mu \vert$
that is exceeded in 1\% of cases,
\begin{equation}
\mu_{1s} = 2.61 \sigma = 2.61 \: n_0^{-1/2} \sigma_0.
\label{eq:mu1s}
\end{equation}

Since the statistics of $\lambda$ are somewhat more complicated, we
will infer the approximate value of $\lambda_{1s}$ from the value of
$\mu_{1s}$:
\begin{equation}
\lambda_{1s} \approx \frac{2\mu_{1s}}{T_0} = 5.22 \: \sigma_0 n_0^{-1/2} 
T_0^{-1}.
\label{eq:lambda1s}
\end{equation}

\subsubsection{Long-Period Regime \label{sec:tI-long}}
In this regime, the analytical expressions for ${\cal A}_c$,${\cal A}_s$, $\lambda$,
and $\mu$ that we derived above no longer hold, and so we resort
to simulations. To this end we create a synthetic data set consisting
of Gaussian noise with $\sigma_0$ = 100 $\mu$as (we simulate this
measurement accuracy because future instruments such as FAME and
Keck Interferometer will probably attain similar accuracies;
Horner et al. 1999; van Belle et al. 1998).  We
sample the synthetic data at one month intervals for $T_0 = 12$
years.  We explore orbital periods from 5 years to 100 years,
choosing the interval between sampled periods so as to result in
a one radian decrease in the total number of orbital cycles
over the length of the observations,
\begin{equation}
\Delta \tau = \frac{\tau^2}{2\pi T_0}.
\label{eq:periods}
\end{equation}
For each period, $\tau$, we simulate $N=1000$ data sets and carry out the
least-squares fit.  We set $A_{c1}$ equal to the 10$^{\rm th}$ highest
value that arises.  Clearly, 99\% of fitted ${\cal A}_c$ will lie below
this value.  Using the same method, we also determine the 99\%
confidence level for the other fitted parameters, which we denote
by $A_{s1}$, $\lambda_{1}$, and $\mu_1$.

\begin{figure}
\plotone{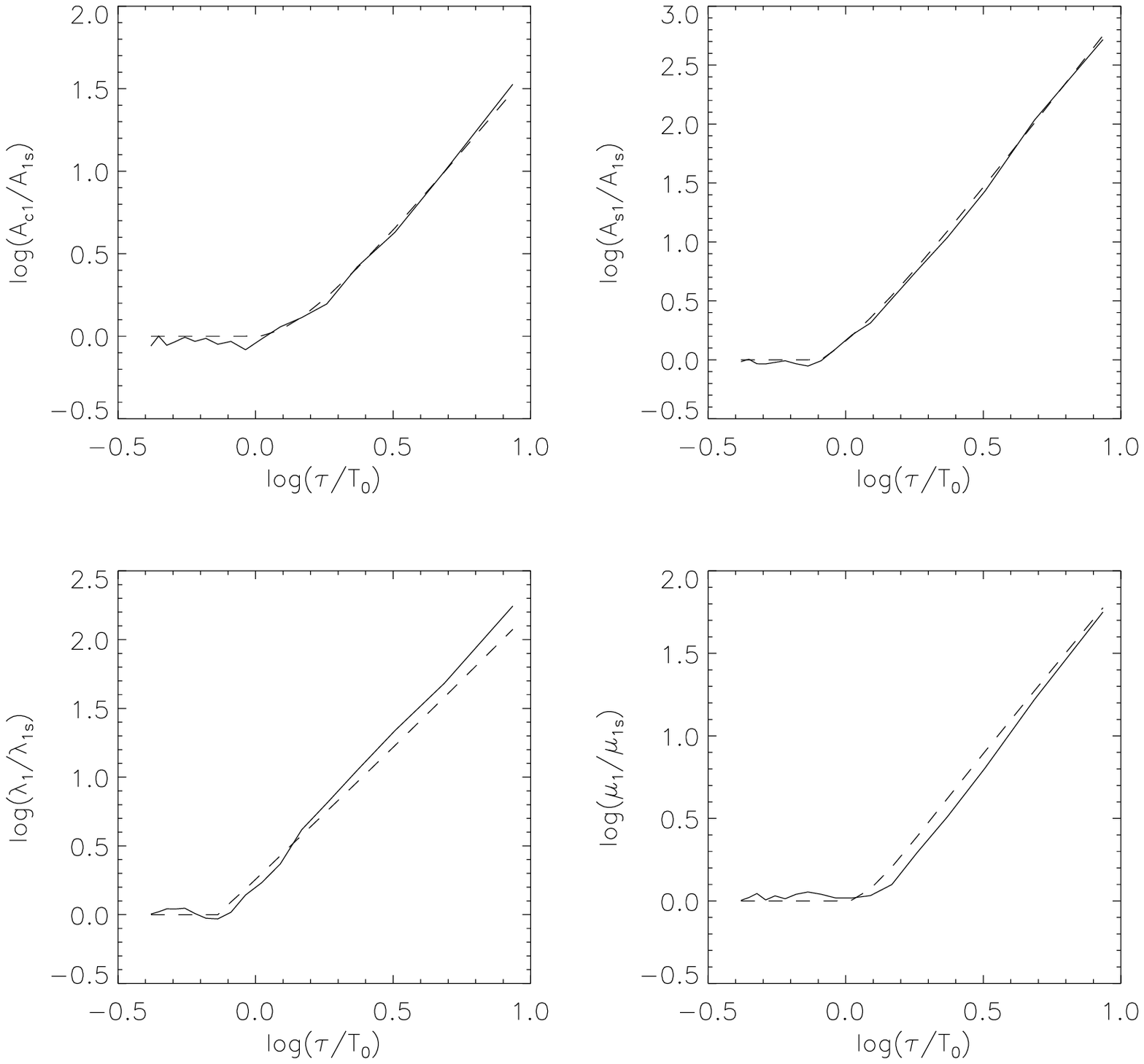}
\caption{Illustration of how the fitted parameters depend on
orbital period in the case of edge-on orbits.  The solid lines indicate
the $99^{\rm th}$ percentile values of the orbital parameters ${\cal A}_c$,
${\cal A}_s$, $\lambda$, and $\mu$.  Here, we simulated 
$N=1000$ data sets with Gaussian noise of zero mean and $\sigma_0 = 100$
$\mu$as.  Data were sampled at one-month intervals for $T_0 = 12$ years.  
The dashed lines show the analytic estimates predicted by
Equations \ref{eq:as}--\ref{eq:mu}.  
\label{fig:params-tau}}
\end{figure}

\begin{figure}
\plotone{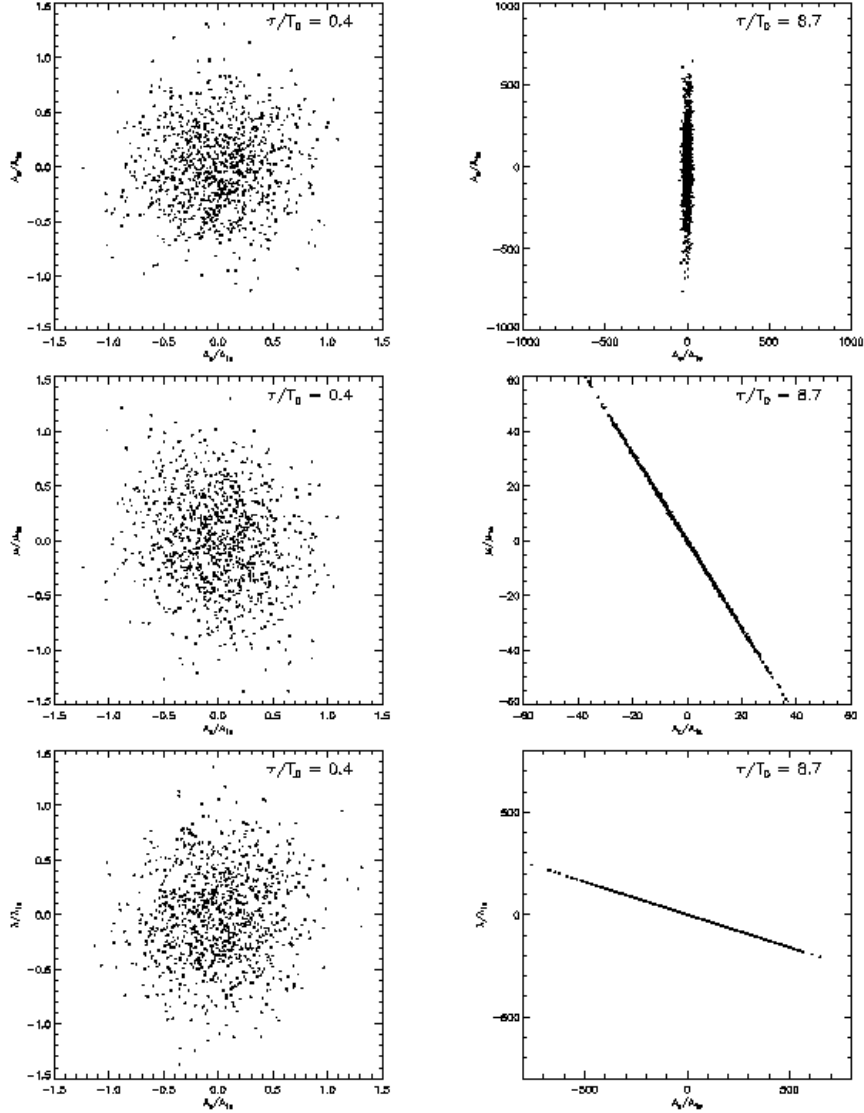}
\caption{Correlations between fitted parameters for edge-on orbits in the 
short- and long-period regimes.  Each dot represents one out of the $N=1000$ 
simulations.  The duration of the astrometric
monitoring is $T_0=12$ yr.  In the short-period regime, $\tau \ll T_0$,
all of the orbital parameters are uncorrelated.  In the long-period regime,
strong correlations are seen.  In particular, orbital phases of
approximately $0^{\circ}$ 
are preferred ({i.e.} the Type I error ellipse has a large
aspect ratio), ${\cal A}_c$ and $\mu$ are anticorrelated,
and ${\cal A}_s$ and $\lambda$ are anti-correlated.
\label{fig:params-corr}}
\end{figure}

We find that the values of the fitted parameters are in excellent
agreement with the values predicted by Equations 
\ref{eq:ac-short}--\ref{eq:mu-short}
in the short-period regime (Figure \ref{fig:params-tau}).
In order to understand the behaviors of the fitted parameters
in the long-period regime, we begin by examining how the covariance
of the fitted parameters depends on period.  We expect that in the short-period
regime, all four parameters will be completely uncorrelated, since
we observe many orbital periods and can thus easily distinguish
the sinusoidal motion from the linear proper motion term and the constant
offset term.  This expectation is verified by the simulations
(Figure \ref{fig:params-corr}).

In the long-period regime, in contrast, strong correlations may arise
between the fitted parameters.  From Figure \ref{fig:params-corr},
we conclude the following for the long-period regime:

\noindent 1. ${\cal A}_c$ and ${\cal A}_s$ are correlated, and select orbital phases 
close to 0$^{\circ}$.

\noindent 2. ${\cal A}_c$ and $\mu$ are anticorrelated.

\noindent 3. ${\cal A}_s$ and $\lambda$ are anticorrelated.

\noindent 4. ${\cal A}_c$ and $\lambda$ are uncorrelated (not shown in figure).

\noindent 5. ${\cal A}_s$ and $\mu$ are uncorrelated (not shown in figure).

\noindent Previous authors noted that these correlations would arise, and
that the orbital phase would become important in the
long-period regime \citep{BS82}.  However,
neither they nor others have explored the full-implications of
this fact.  Below, we look into this issue in more detail, and
develop an analytical understanding of how the fitted 
parameters depend upon orbital period.  

We now provide a physical explanation of the results
displayed in Figure \ref{fig:params-corr}.  When $\tau \ll T_0$,
sinusoids with random phase can be fit to Gaussian data,
where the amplitude of the sinusoids is constrained by the
vertical scale of the data ($\sim \sigma_0$).  However,
when $\tau > T_0$, this is no longer necessarily the case.
The maximum amplitude is obtained when a sine wave centered
around $t=0$ is fitted.  This is because for $\tau > T_0$,
$\sin(\pi T_0 / \tau) \approx \pi T_0/ \tau$, which can be
completely absorbed into the linear proper motion term.

Specifically, the maximum amplitude of the sine wave that
can be fitted is limited by how much $\sin(2\pi t/\tau)$ deviates from
a straight line in the range from $t=[0,T_0/2]$.  This
deviation is given by
\begin{equation}
\frac{\pi T_0}{\tau} - \sin\left(\frac{\pi T_0}{\tau}\right).
\label{eq:sin-dev}
\end{equation}
We note that the maximal deviation occurs when $\tau \ge 3T_0/4$,
and thus the long-period regime ``begins'' where $\tau=3T_0/4$.
The largest possible amplitude of the fitted sine wave (${\cal A}_s$) 
in the long-period regime is that for which
this deviation is approximately equal to the vertical scale of the
noise ($\sim \sigma_0$).  Since Equation \ref{eq:sin-dev}
tells us the fractional observed amplitude, the real
amplitude of the fitted sine is
\begin{equation}
A_{s1} = \frac{A_{1s} [4\pi/3 - \sin(4\pi/3)]}{\pi T_0 / \tau
- \sin\left(\pi T_0/ \tau\right)}.
\label{eq:as}
\end{equation}
Here, $A_{s1}$ is the value of $\vert {\cal A}_s \vert$ that is exceeded
in 1\% of least-squares fits to Gaussian noise, $A_{1s}$ is the
corresponding value of $A_{s1}$ in the short-period regime
(Equation \ref{eq:a1s}), and the
specific normalization is chosen so that at the boundary of the short
and long-period regimes ($t=3T_0 / 4$), $A_{s1}$ is continuous.

We can understand the behavior of ${\cal A}_c$ in the long-period regime using
a similar analysis.  Specifically, the maximum amplitude of a
cosine wave centered around $t=0$ is constrained by how much
this cosine deviates from a constant in the interval $[0,T_0/2]$.
Following the same lines as the analysis for ${\cal A}_s$, we find that
this amplitude is given by
\begin{equation}
A_{c1} = \frac{2A_{1s}}{1-\cos\left({\pi T_0/\tau}\right)}.
\label{eq:ac}
\end{equation}
Here, $A_{c1}$ is the value of $\vert {\cal A}_c \vert$ that is exceeded
in 1\% of least-squares fits to Gaussian noise, $A_{1s}$ is the
corresponding value of $A_{c1}$ in the short-period regime
(Equation \ref{eq:a1s}), and 2 is a normalization factor (chosen to
enforce continuity between the short and long-period regimes).
We note that in the long-period regime $\sin(x) \rightarrow x$ faster than 
$\cos(x) \rightarrow 1$, and thus the covariance
of the sinusoidal signal with the constant offset term is less
significant than the covariance with the proper motion
term.

In light of these analytical expressions for ${\cal A}_c$ and ${\cal A}_s$, it is
not difficult to find expressions for $\lambda$ and $\mu$.  In
particular, because the input data consists of Gaussian noise with a
mean of zero, we know that $\lambda$ and $\mu$ must be chosen 
so that the resultant fit is also, in the mean, zero.  
The anti-correlation of ${\cal A}_c$ and $\mu$ and that of ${\cal A}_s$ and
$\lambda$ confirm this expectation.  

We must choose a $\lambda$ that ``cancels'' the fitted sine-wave
component.  In the long-period regime, this condition can be
expressed as ${\cal A}_s \sin(\omega t) \approx {\cal A}_s \omega t \sim \lambda t$,
or $\lambda \sim {\cal A}_s / \tau$.  However, because the data has some vertical
scale, we must vertically ``re-center'' the data after subtracting
${\cal A}_s/\tau$.  Since the total vertical scale is $\sim 2\sigma_0$,
we must add $\sigma_0$ after subtracting ${\cal A}_s/\tau$.  Thus,
we expect that for $\tau \ge 3T_0/4$,
\begin{equation}
\lambda_{1} = \lambda_{1s} \frac{3 T_0}{4\tau} 
\left\{2 \frac{{4\pi}/{3} -
\sin\left({4\pi}/{3}\right)}{{\pi T_0}/{\tau}
-\sin\left({\pi T_0}/{\tau}\right)} - 1\right\}.
\label{eq:lambda}
\end{equation}

Similarly, $\mu$ is chosen to ``cancel'' the cosine component.
Thus, $\mu$ is the negative cosine component
(given by ${\cal A}_c$) plus $\sigma_0$ (where $\sigma_0$ is needed
to reproduce the data, which is vertically centered around 0).
Thus, for $\tau \ge T_0$,
\begin{equation}
\mu_1 = \mu_{1s} \left[\frac{4}{1-\cos\left({\pi T_0}/{\tau}\right)}
-1\right].
\label{eq:mu}
\end{equation}

Equipped with the behaviors of $A_{c1}$ and $A_{s1}$, we now know
the analytical behavior of the Type I errors in the long-period
regime.  Specifically, the error ellipse within which 99\% of fitted
signals lie, $\epsilon_1$, is completely described:
$A_{s1}$ gives the semi-major axis, and $A_{c1}$ gives
the semi-minor axis. 

\noindent{\bf Amplitude--Phase Analysis.}
The standard signal analysis usually focuses on the statistics of
the orbital amplitude ${\cal A} = 
({\cal A}_c^2 + {\cal A}_s^2)^{1/2}$.  This is
adequate when ${\cal A}_c$ and ${\cal A}_s$ are uncorrelated ({e.g.} in the
short-period regime with adequate sampling).  However, as can
be seen in Figure \ref{fig:params-corr}, the ${\cal A}_c$--${\cal A}_s$
distribution becomes highly elliptical in the long-period
regime.  Thus in this regime it pays to undertake an analysis
that utilizes amplitude and phase.
The fact that the amplitude-phase ellipse (Figure
\ref{fig:params-corr}) is significantly
smaller than the amplitude-only circle in the long-period
regime clearly indicates the superiority of an
amplitude-phase approach.  Therefore, we will use an
amplitude-phase approach in the remainder of this discussion.
See EK2001 for a more detailed
comparison of amplitude-only and amplitude-phase approaches.

\subsection{Type II Errors \label{sec:typeII}}
In \S \ref{sec:typeI}, we computed the probability of detecting an
apparent signal generated purely by noise ({i.e.} Type I probabilities).
We now consider Type II probabilities-- the probability of failing
to detect a genuine signal due to contamination by noise.  
To this end, we simulate a data set that consists of signal and noise:
\begin{equation}
\theta'(t_i) = {\cal A} \sin(\omega t_i + \phi) + N(t_i),
\label{eq:signal}
\end{equation}
where ${\cal A}$ is the amplitude of the signal, $\phi$ is the phase, and
$N(t_i)$ is the Gaussian noise.  We let $\phi$ be drawn from a 
uniform distribution in the interval $[0,2\pi]$, an appropriate
assumption for circular orbits.  We choose an initial signal
amplitude of $\sigma_0/2$, and then do $N=1000$ least-squares
fits (with the same parameters as in \S \ref{sec:typeI}).

\begin{figure}
\plotone{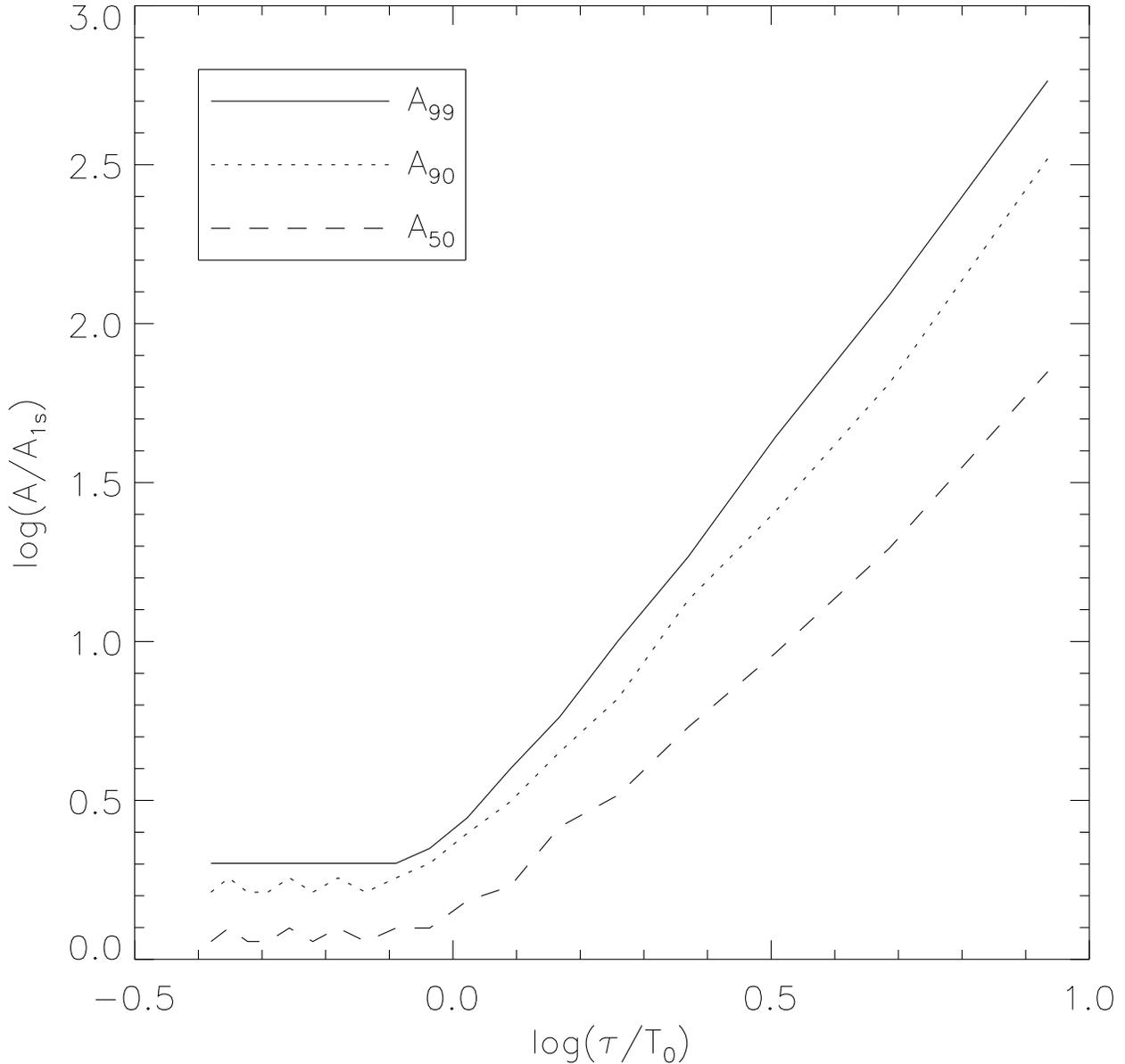}
\caption{Plots of $A_{99}$, $A_{90}$, and $A_{50}$ and a function of
orbital period, for edge-on orbits.  
$A_{99}$ is the signal amplitude necessary such that
least-squares fits to data containing a genuine signal  
plus Gaussian noise will yield detections 99\% of the time.
We simulated $N=1000$ data sets sampled at one-month intervals
for $T_0 = 12$ years.  The data sets consist of Gaussian noise (with
zero mean and $\sigma_0 = 100$ $\mu$as) plus signal;
the phase of the signal was assumed to be randomly and evenly
distributed over the range $[0,2\pi]$. 
\label{fig:edge-k99}}
\end{figure}

In \S \ref{sec:typeI}, we showed that in the long-period regime, the
${\cal A}_c$--${\cal A}_s$ distribution is elliptical.  
In view of this, we define
the Type I errors by an ellipse $\epsilon_1$; this ellipse contains
99\% of the zero-signal simulations (\S \ref{sec:tI-long}).
In our analysis of Type II probabilities, we
consider all least-squares fits (of the signal specified in Equation 
\ref{eq:signal} to the model specified by Equation \ref{eq:lsq-model1})
that lie within $\epsilon_1$ to be indistinguishable from those
produced by noise.  We increment the input signal 
amplitude until a certain fraction of fitted points lie outside
of $\epsilon_1$.  For example, the signal amplitude at which 99\%
of fitted points lie outside of $\epsilon_1$ is denoted by $A_{99}$.
Plots of $A_{99}$, $A_{90}$, and $A_{50}$ (as a function of
orbital period) are shown in Figure \ref{fig:edge-k99}.

\section{Face-On Orbits \label{sec:face}}
In the case of face-on orbits, we have twice as many measurements
as in the edge-on case, since there are two orbital dimensions.  Thus,
our model will contain both the $x$ and $y$ motion of the orbit:
\begin{equation}
\theta_x(t) = {\cal A}_c \cos(\omega t) + {\cal A}_s \sin(\omega t) + \lambda_x t + \mu_x,
\label{eq:lsq-modelx}
\end{equation}
\begin{equation}
\theta_y(t) = -{\cal A}_s \cos(\omega t) + {\cal A}_c \sin(\omega t) + \lambda_y t + \mu_y.
\label{eq:lsq-modely}
\end{equation}
As above, ${\cal A}_c={\cal A}\sin\phi$, ${\cal A}_s={\cal A}\cos\phi$, $\vec{\lambda}=[\lambda_x,
\lambda_y]$ is the proper motion, 
and $\vec{\mu}=[\mu_x,\mu_y]$ is the position offset.

Using the same parameters as in \S \ref{sec:lsq}, we determine the
unknown model parameters by minimizing
\[X^2 = \sum_{j=0}^{n_0-1} \{[\theta_x'(t_j)-{\cal A}_c\cos(\omega t_j) + 
{\cal A}_s\sin(\omega t_j)-\lambda_x t_j - \mu_x]^2 + \]
\begin{equation}
[\theta_y'(t_j) +
{\cal A}_s\cos(\omega t_j)-{\cal A}_c\sin(\omega t_j) - \lambda_y t_j - \mu_y]^2\}
\label{eq:x2-face}
\end{equation}
with respect to ${\cal A}_c$, ${\cal A}_s$, $\lambda_x$, $\lambda_y$, $\mu_x$, and $\mu_y$.

\subsection{Type I Errors \label{sec:tI-face}}
Because a face-on orbit has two observable dimensions, the analysis
of Type I errors will be somewhat more complicated than for the face-on case.
First, we note that in the short-period regime where $\tau < T_0$,
the least squares fit essentially has twice as many measurements
to work with.  Thus, we expect that $A_{1s}$ will be $2^{-1/2}$ times
its value for the edge-on case:
\begin{equation}
A_{1s} = 2.61 \sigma = 2^{-1/2}\times 3.69 \: n_0^{-1/2} \sigma_0.
\label{eq:a1s-face}
\end{equation}
However, since ${\lambda}$ and ${\mu}$ must be determined for each
orbital dimension independently, $\mu_{1s}$ and $\lambda_{1s}$ will
still be given by Equations \ref{eq:mu1s} and \ref{eq:lambda1s}.

When $\tau>T_0$, the full amplitude of a cosine centered around $t=0$
cannot be observed.  However, in the case of face-on orbits, we
can observe the full amplitude of the orbit in the orthogonal
direction, since it is phase-shifted by 90$^{\circ}$.  
The full amplitude of this sine wave
can be observed until $\tau>4T_0/3$ (to see this, find the value
of $\tau$ for which $\cos(\pi T_0/\tau) = -\sin(\pi T_0/\tau)$).

For face-on orbits, the long-period regime begins when $\tau>4T_0/3$,
since it is here that the full amplitude of a sinusoid cannot be
observed in either orbital dimension.  Moreover, since the two dimensions
are phase-shifted by $90^{\circ}$, we expect the expressions for $A_{c1}$
and $A_{s1}$ to be the same.  Since $\sin(x) \rightarrow x$ faster than
$\cos(x) \rightarrow 1$, the deviation of a cosine wave from unity
will be the important constraint on the possible fitted amplitude.  
Thus, $A_{c1}$ and $A_{s1}$ are both given by Equation \ref{eq:ac} in
the long-period regime.  

\begin{figure}
\plotone{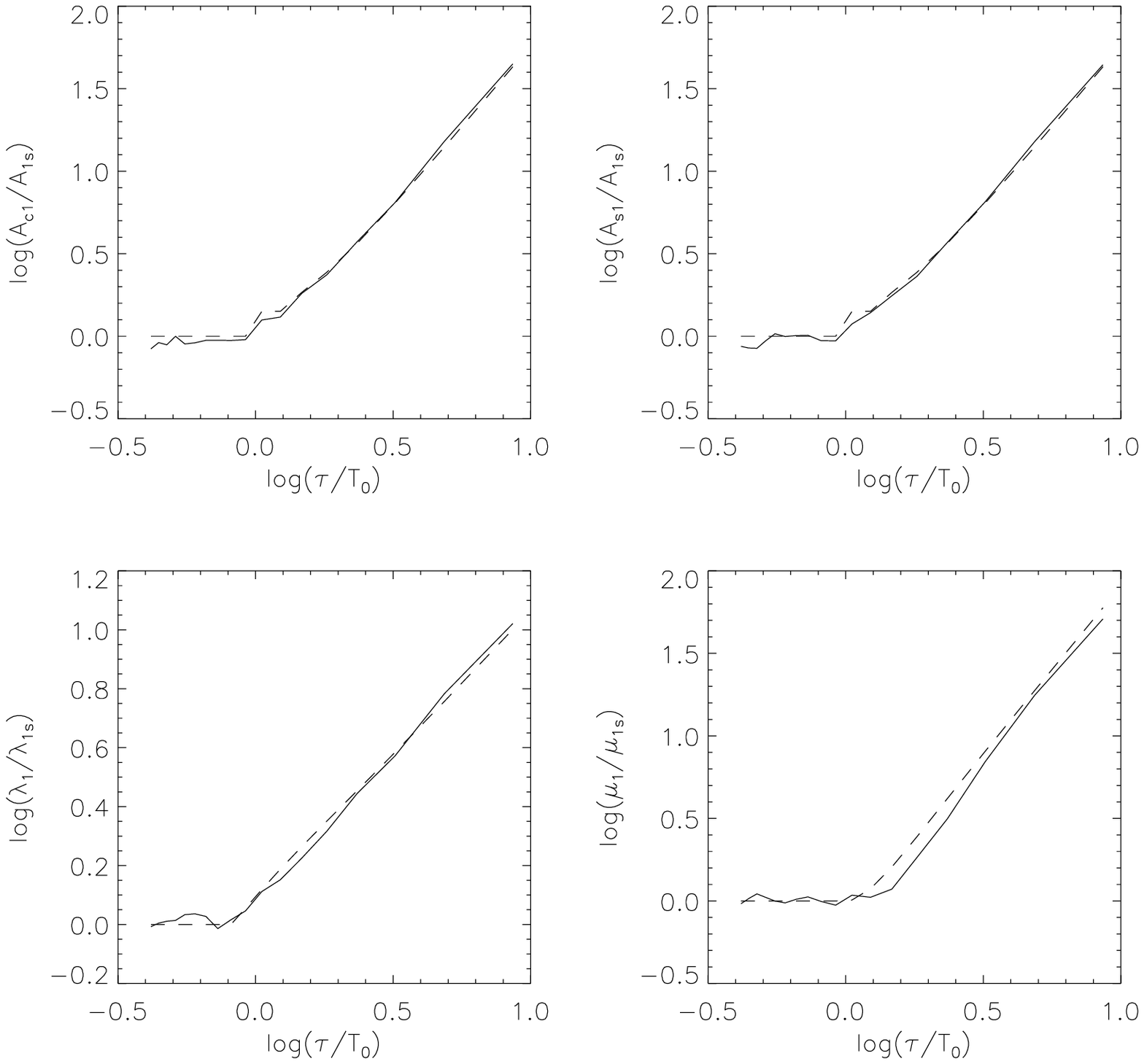}
\caption{Illustration of how the fitted parameters depend on
orbital period in the case of face-on orbits. The solid lines indicate
the $99^{\rm th}$ percentile values of the orbital parameters ${\cal A}_c$,
${\cal A}_s$, $\lambda$, and $\mu$, and the
dashed lines show the analytic estimates predicted in \S \ref{sec:tI-face}.
See Figure \ref{fig:params-tau} for details of the simulations.
\label{fig:params-tau-face}}
\end{figure}

A potential complication is that at some point,
we are gaining virtually no new information from one of the orbital
dimensions (since the constraint on the size of a cosine is so
much stronger than that on the size of a sine wave).  So, we expect that
at some period, our sensitivity will drop roughly by a factor of $2^{1/2}$,
to reflect the fact that half of the measurements become useless.
We model this behavior as follows:
\begin{equation}
A_{s1} = A_{c1} = \cases{A_{1s} & for $\tau < T_0$ \cr \sqrt{2}A_{1s} & for
$T_0 < \tau < 4T_0/3$ \cr
\frac{2\sqrt{2}A_{1s}}{1-\cos\left({\pi T_0}/{\tau}\right)} & for
$\tau > 4T_0/3$ \cr}
\label{eq:ac-face}
\end{equation}
As illustrated by Figure \ref{fig:params-tau-face}, this function is
a good fit to the simulations.

Since $A_{s1}$ now depends on the maximum amplitude of a cosine wave,
rather than the amplitude of a sine wave, we must modify our expression
for $\lambda_1$ accordingly.  Specifically, while the slope of a sine
wave centered around $t=0$ goes approximately as $\omega t$, the
slope of a cosine wave goes as $\omega ^2 t^2$.  Thus, we expect that
the expression for $\lambda_1$ in the case of face-on orbits will 
differ from that for edge-on orbits by a factor of $\omega$.  Specifically, 
we find that for $\tau > T_0$,
\begin{equation}
\lambda_{1} = \lambda_{1s} \left(\frac{T_0}{\tau}\right)^2
\left\{2 \frac{\pi - \sin\pi}{{\pi T_0}/{\tau}
-\sin\left({\pi T_0}/{\tau}\right)} - 1\right\}.
\label{eq:lambda-face}
\end{equation}
This analytic function is shown along with the simulated data
in Figure \ref{fig:params-tau-face}.

We expect that $\mu_1$ will be the exactly the same as in the case
of edge-on orbits, since the behavior of $A_{c1}$ is unchanged.  This
expectation is confirmed by Figure \ref{fig:params-tau-face}.

\begin{figure}
\plotone{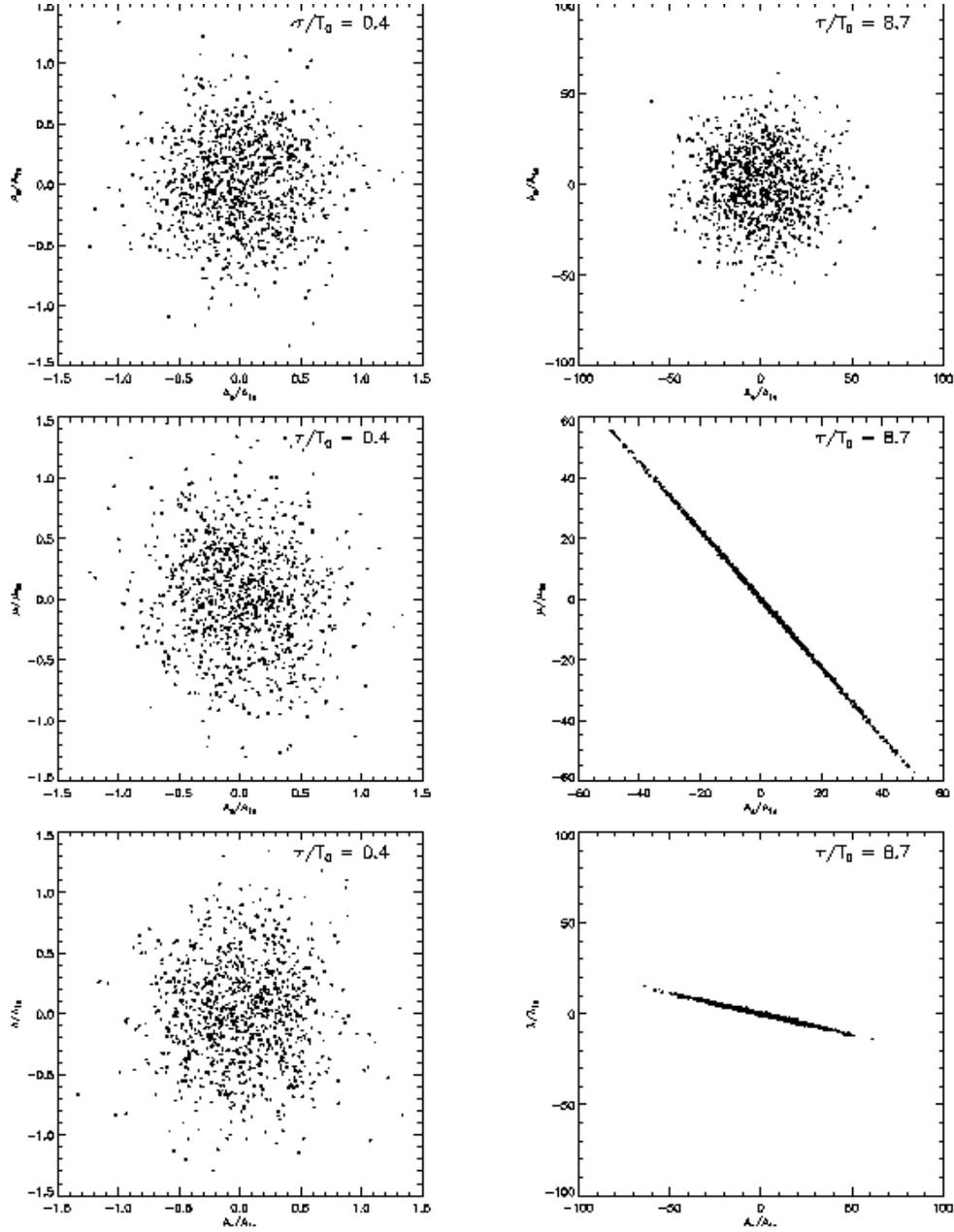}
\caption{Correlations between fitted parameters for face-on orbits.
Each dot represents one out of the $N=1000$ 
simulations.  The duration of the astrometric
monitoring is $T_0=12$ yr.  In the case of face-on orbits, we see that
${\cal A}_c$ and ${\cal A}_s$ remain uncorrelated in the long-period regime, and thus
the aspect ratio of the Type I error ellipse $\epsilon_1$ is unity.
\label{fig:params-corr-face}}
\end{figure}

From these analytic expressions for ${\cal A}_c$ and ${\cal A}_s$, we see that
the sensitivity of astrometric technique to face-on orbits is
significantly better than that for edge-on orbits.  This is well
illustrated through a comparison of Figures \ref{fig:params-corr} and
\ref{fig:params-corr-face}.  As discussed above, the Type I error ellipse,
$\epsilon_1$, is defined as the ellipse in the ${\cal A}_c$--${\cal A}_s$ 
plane for which 99\%
of fitted parameters lie inside.  In the case of edge-on orbits, this
ellipse is very eccentric, since the semi-major axis is given by Equation
\ref{eq:as} and the semi-minor axis is given by Equation \ref{eq:ac}
(Figure \ref{fig:params-corr}).  In contrast, in the case of face-on orbits,
$\epsilon_1$ is a circle whose radius is given by Equation \ref{eq:ac-face}.
Thus, for face-on orbits, we are equally sensitive to signals of arbitrary
orbital phase.

\subsection{Type II Errors \label{sec:tII-face}}
As in \S \ref{sec:typeII}, we compute the probability that a real signal 
will not be detected due to contamination by noise. For a face-on orbit, 
we simulate the following data set:
\begin{equation}
\theta'_x(t_i) = {\cal A}\sin(\omega t_i + \phi) + N_x(t_i), 
\label{eq:face-sigx}
\end{equation}
\begin{equation}
\theta'_y(t_i) = {\cal A}\cos(\omega t_i + \phi) + N_y(t_i), 
\label{eq:face-sigy}
\end{equation}
Here, ${\cal A}$ is the amplitude of the signal, $\phi$ is the 
(randomly-distributed) orbital phase, and $N(t_i)$ is the Gaussian noise. 
Since the sensitivity of astrometric technique to face-on orbits is better 
than for edge-on orbits by a factor of $2^{1/2}$ (in the short-period regime),
we choose an initial signal amplitude of $\sigma_0/2^{3/2}$. We then do 
$N = 1000$ least-squares fits, with the same parameters as in \S
\ref{sec:tI-face}.

\begin{figure}
\plotone{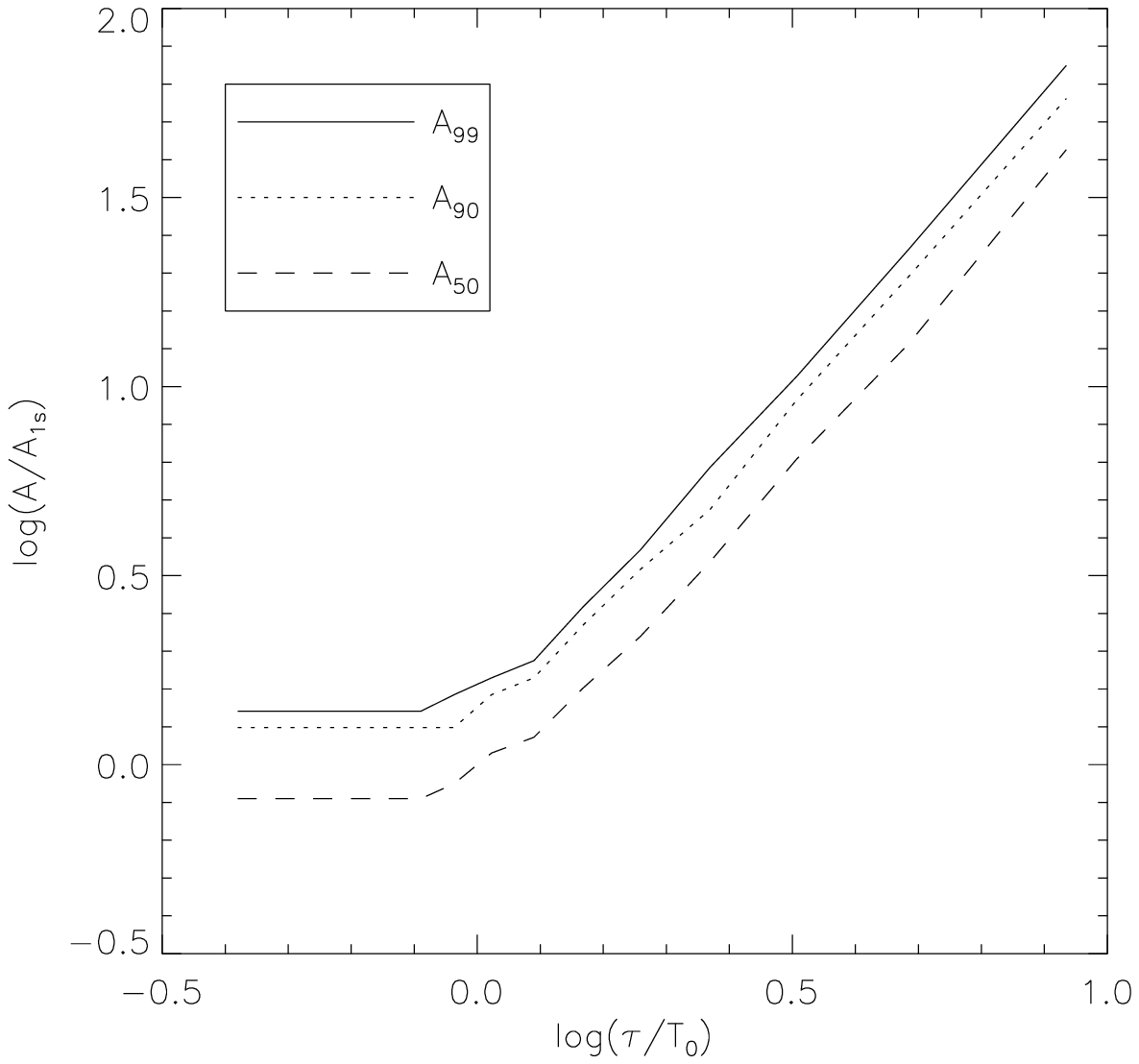}
\caption{Plots of $A_{99}$, $A_{90}$, and $A_{50}$ as a function of
orbital period, in the case of face-on orbits.  $A_{99}$ is the signal 
amplitude necessary such that
least-squares fits to data containing a genuine signal of random
phase plus Gaussian noise will yield detections 99\% of the time.
See Figure \ref{fig:edge-k99} for details of the simulations.
\label{fig:k99-face}}
\end{figure}

Following our analysis of Type II errors in \S \ref{sec:typeII}, we consider 
all least-squares fits (of the signal specified in Equations \ref{eq:face-sigx}
and \ref{eq:face-sigy} to the model given by Equations \ref{eq:lsq-modelx} and 
\ref{eq:lsq-modely}) that lie within $\epsilon_1$ to be indistinguishable 
from noise. The results of this analysis show that the sensitivity of the 
astrometric technique is much better for face-on orbits than for edge-on 
orbits (Figure \ref{fig:k99-face}).

\section{Inclined Orbits \label{sec:45}}
In the case of generally inclined orbits, we observe the
full amplitude in one orbital dimension, but only $\sin i
\times {\cal A}$ in the other dimension (where $i$ is the orbital inclination
angle).  Labeling the dimension in which
we measure the full amplitude by $x$, our model becomes
\begin{equation}
\theta_x(t) = {\cal A}_c \cos(\omega t) + 
{\cal A}_s \sin(\omega t) + \lambda_x t + \mu_x,
\label{eq:lsq-modelx-45}
\end{equation}
\begin{equation}
\theta_y(t) = -\sin i \: {\cal A}_s \cos(\omega t) + \sin i \:
{\cal A}_c \sin(\omega t) + \lambda_y t + \mu_y.
\label{eq:lsq-modely-45}
\end{equation}
As above, ${\cal A}_c={\cal A}\sin\phi$, ${\cal A}_s={\cal A}\cos\phi$, 
$\vec{\lambda}$ is the proper motion, 
and $\vec{\mu}$ is the position offset.

Using the same parameters for our data set as in \S \ref{sec:lsq}, 
we determine the unknown model parameters by minimizing
\[X^2 = \sum_{j=0}^{n_0-1} \{[\theta_x'(t_j)-{\cal A}_c\cos(\omega t_j) - 
{\cal A}_s\sin(\omega t_j)-\lambda_x t_j - \mu_x]^2 + \]
\begin{equation}
[\theta_y'(t_j) +
\sin i \: {\cal A}_s\cos(\omega t_j)- \sin i \: {\cal A}_c\sin(\omega t_j)
- \lambda_y t_j - \mu_y]^2\}
\label{eq:x2-45}
\end{equation}
with respect to ${\cal A}_c$, ${\cal A}_s$, $\lambda_x$, $\lambda_y$, $\mu_x$, and $\mu_y$.

\subsection{Type I Errors \label{sec:tI-45}}
In the short-period regime where $\tau < T_0$, the least squares fit 
essentially has twice as many measurements with which to work, although half 
of the measurements are less sensitive. Thus, we expect that Type I errors 
will be somewhere between those for the edge-on and face-on cases. 
However, for 
simplicity, we will simply use the expressions from the face-on case.

As above, $A_{c1}$ is given by Equation \ref{eq:ac}
in the long-period regime, since we can always observe the full amplitude of 
a cosine wave in one orbital dimension. However, $A_{s1}$ will be different 
than in the face-on case, since the cosine wave cannot be measured as 
accurately in the other orbital dimension. Thus, we expect that in the 
long-period regime, $A_{s1}$ 
will be stretched relative to $A_{c1}$.  For orbits with inclination angles
close to zero, there is essentially no information from the second
orbital dimension, and thus $A_{s1}$ is given by Equation \ref{eq:as}.
However, for inclination angles significantly greater than zero, the
information from the second orbital dimension becomes important,
and $A_{s1}$ is limited by the lower-amplitude cosine wave in this
second dimension.  Specifically, $A_{s1}$ is $A_{c1}$ stretched by
$1/\sin i$.
Thus, we have an expression for the aspect ratio of $\epsilon_1$
as a function of inclination angle:
\begin{equation}
\frac{A_{s1}}{A_{c1}} = \min\left(\frac{1}{\sin i}, 
\left[\frac{A_{c1}}{A_{s1}}\right]_{\rm edge}\right).
\label{eq:as1-45}
\end{equation}

\begin{figure}
\plotone{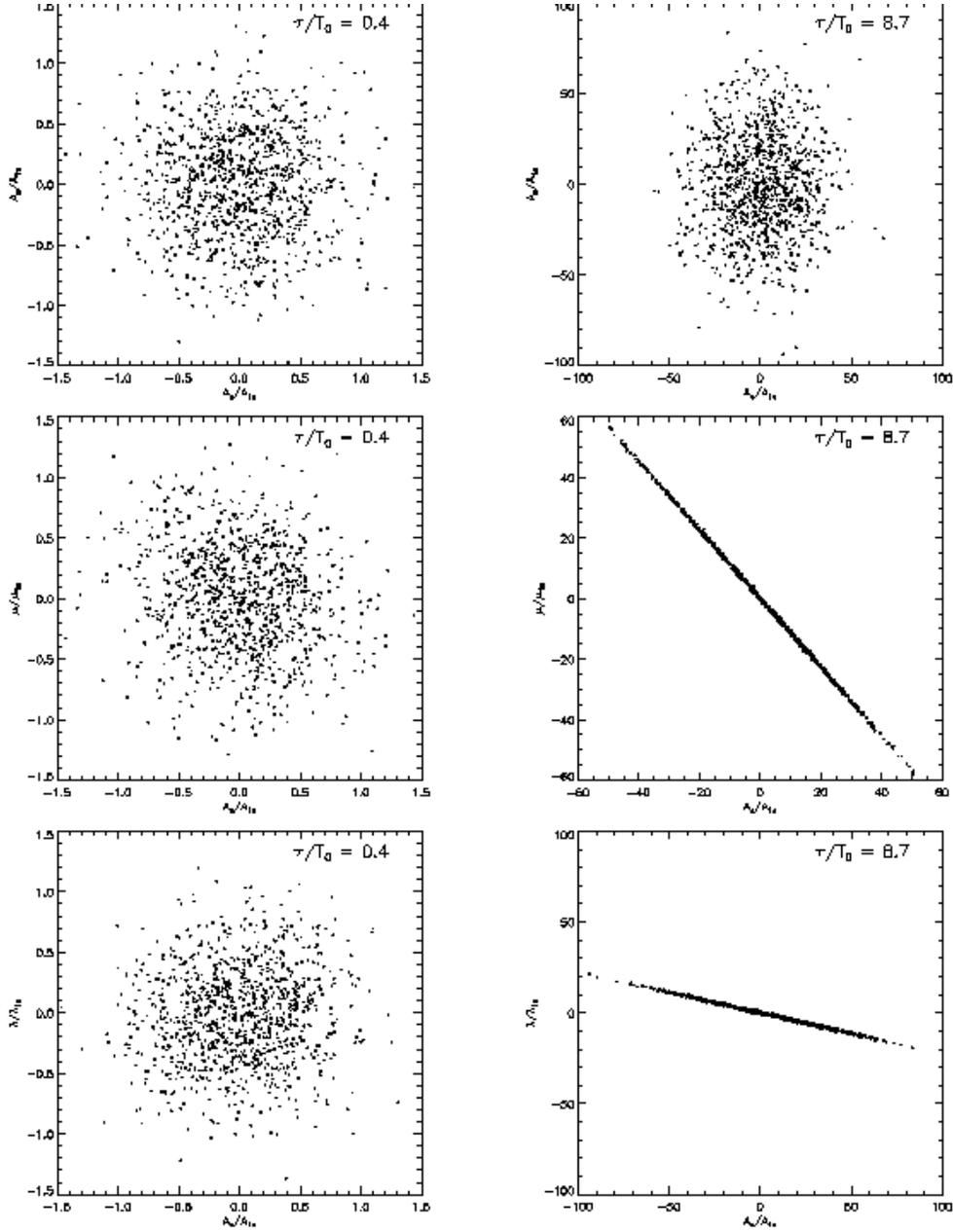}
\caption{Correlations between fitted parameters for  $45^{\circ}$--inclined
orbits.  Each dot represents one out of the $N=1000$ 
simulations.  The duration of the astrometric
monitoring is $T_0=12$ yr.  In the case of $45^{\circ}$--inclined orbits, 
we see that
the aspect ratio of the Type I error ellipse $\epsilon_1$ is $2^{1/2}$
in the long-period regime.
\label{fig:params-corr-45}}
\end{figure}

\begin{figure}
\plotone{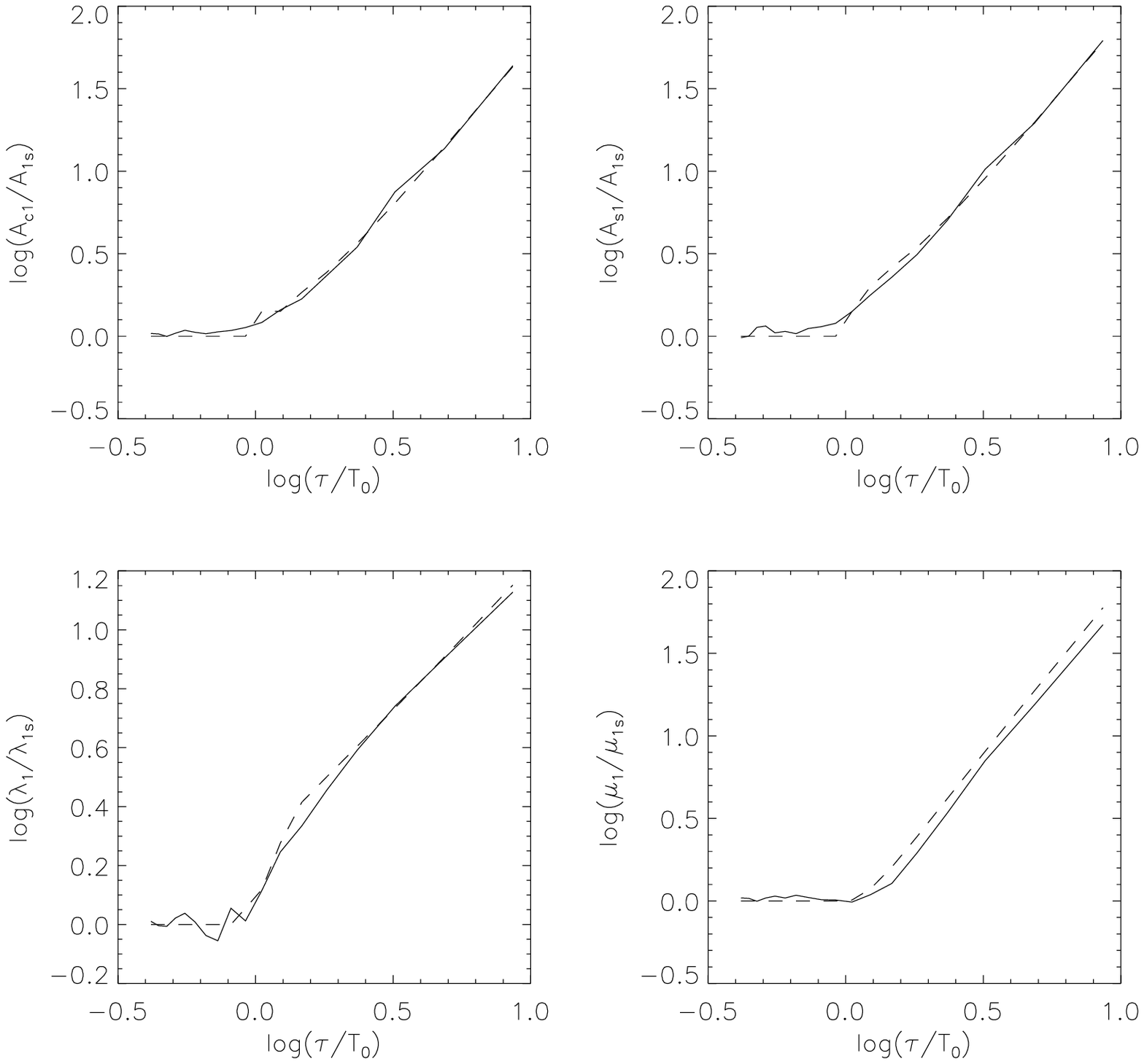}
\caption{Illustration of how the fitted parameters depend on
orbital period in the case of $45^{\circ}$--inclined orbits.
The solid lines indicate
the $99^{\rm th}$ percentile values of the orbital parameters ${\cal A}_c$,
${\cal A}_s$, $\lambda$, and $\mu$, and the
dashed lines show the analytic estimates predicted in \S \ref{sec:tI-45}.
See Figure \ref{fig:params-tau} for details of the simulations.
\label{fig:params-tau-45}}
\end{figure}

As an example, we determine the Type I errors for a $45^{\circ}$--inclined
orbit.  In this case, $A_{s1} = A_{c1} / \sin i = 2^{1/2} A_{c1}$
(Figure \ref{fig:params-corr-45}).
Since $A_{s1}$ is increased by $2^{1/2}$ from the face-on case, $\lambda_1$
must be similarly increased:
\begin{equation}
\lambda_{1}= \sqrt{2} \lambda_{1s} \left(\frac{T_0}{\tau}\right)^2
\left\{2 \frac{\pi-\sin\pi}{{\pi T_0}/{\tau} - \sin({\pi T_0}/{\tau})}
-1\right\}.
\label{eq:lambda1-45}
\end{equation}
We expect that $\mu_1$ will be the exactly the same as in the case of 
edge-on orbits, since the behavior of $A_{c1}$ is unchanged. 
Analytic functions for the parameters are shown along with the simulated 
data in Figure \ref{fig:params-tau-45}.

\subsection{Type II Errors} 
As in \S \ref{sec:typeII} we compute the probability that a real signal will 
not be detected due to contamination by noise. For inclined orbits, we 
simulate the following data set:
\begin{equation}
\theta'_x(t_i) = {\cal A}\sin(\omega t_i + \phi) + N_x(t_i), 
\label{eq:face-sigx-45}
\end{equation}
\begin{equation}
\theta'_y(t_i) = \sin i \: {\cal A}\cos(\omega t_i + \phi) + N_y(t_i), 
\label{eq:face-sigy-45}
\end{equation}
Here, ${\cal A}$ is the amplitude of the signal, $\phi$ is the 
(randomly-distributed) orbital phase, and $N(t_i)$ is the Gaussian noise. 
We choose an initial signal amplitude, and do 
$N = 1000$ least-squares fits, with the same parameters as in \S
\ref{sec:tII-face}.  

\begin{figure}
\plotone{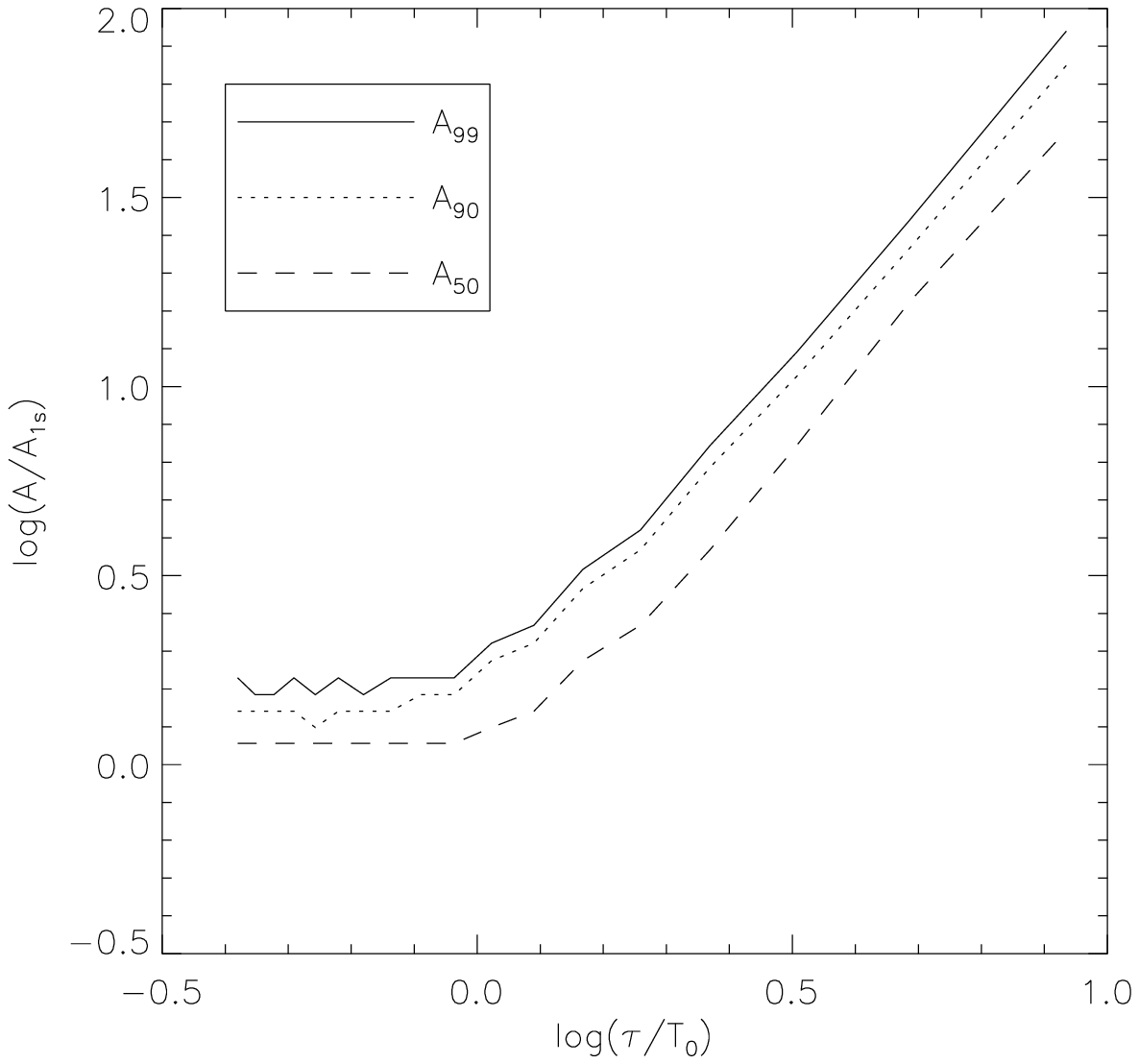}
\caption{Plots of $A_{99}$, $A_{90}$, and $A_{50}$ as a function of
orbital period, in the case of $45^{\circ}$--inclined orbits.  
$A_{99}$ is the signal amplitude necessary such that
least-squares fits to data containing a genuine signal of random
phase plus Gaussian noise will yield detections 99\% of the time.
See Figure \ref{fig:edge-k99} for details of the simulations.
\label{fig:k99-45}}
\end{figure}

Following our analysis of Type II errors in \S \ref{sec:typeII}, we consider 
all least-squares fits (of the signal specified in Equations 
\ref{eq:face-sigx-45} and \ref{eq:face-sigy-45} to 
the model given by Equations \ref{eq:lsq-modelx-45} and 
\ref{eq:lsq-modely-45})
that lie within $\epsilon_1$ to be indistinguishable from noise. 
The results of this analysis for a  $45^{\circ}$--inclined orbit are shown 
in Figure \ref{fig:k99-45}.

\section{Caveats and Conclusions \label{sec:concls}} 
As in our earlier analysis of the sensitivity of the radial velocity 
technique (EK2001), the analysis presented here is based on frequentist 
statistics. However, the importance of the Bayesian approach is gradually 
being recognized in astronomy. Berger \& Delampady (1987) point out that the 
probability of extreme events predicted by the frequentist approach (which is 
what we have essentially computed) is rosier than is warranted. Thus, 
given a mix of stars (with and without planets), the frequentist tail 
criterion of say 1\% (e.g. $\epsilon_1$) is likely to underestimate the 
fraction of false detections that are expected to occur in
population analyses. These criticisms notwithstanding, we have pursued the 
frequentist approach because of (1) its conceptual simplicity, (2) its ease 
of use, (3) our own familiarity, and (4) our lack of familiarity with 
the Bayesian approach.

We also acknowledge that out treatment of the statistics was not always 
completely rigorous. Specifically, when one orbital dimension becomes 
much more sensitive than the other, we simply ignore the data from the 
other dimension. Inclusion of this less sensitive dimension in the analysis 
is likely to lead to a small improvement in the overall sensitivity. We feel 
that this statistical analysis is beyond the scope of this paper, and will 
not lead to any meaningful modifications of the results derived above.

A final caveat is that in this analysis, we have assumed that the orbital 
inclination angle is known {\em a priori}. In general, this is not the case, 
and the inclination angle must be determined simultaneously with the rest of 
the orbital parameters (using a non-linear least-squares fitting method). 
While one expects that all of the sensitivities in this paper are slightly 
under-estimated as a result, it is clear that the qualitative behavior of the 
sensitivity in the long-period regime is unaffected.

Despite the fact that our analysis can be improved upon in some areas, this 
work provides an accurate analytic description of the sensitivity of the 
astrometric technique in detecting outer planets. Moreover, we have 
understood, both qualitatively and quantitatively, how the sensitivity 
improves as the orbital inclination angle is increased.  Of specific
importance is the fact that
in the long-period regime the sensitivity of astrometric technique to
face-on orbits is more that $2^{1/2}$ times better the sensitivity to
edge-on orbits, due to the orthogonal information supplied by the two
orbital dimensions.

\epsscale{0.8}

\epsscale{1.0}

\epsscale{0.8}

\epsscale{1.0}

\epsscale{0.8}

\epsscale{1.0}

\end{document}